# Phénomènes de vieillissement, rajeunissement et mémoire : l'exemple des verres de spin

**Vincent Dupuis — Eric Vincent — Fabrice Bert — Didier Hérisson — Jacques Hammann — Miguel Ocio**

*Service de Physique de l'Etat Condensé (SPEC)*
*CEA Saclay, DSM/DRECAM/SPEC*
*91191 Gif sur Yvette Cedex, France*
*E-mail. : vincent@drecam.saclay.cea.fr*

*RÉSUMÉ. Les verres de spin, systèmes magnétiques désordonnés et frustrés, montrent des phénomènes de vieillissement qui sont aussi caractéristiques des verres structuraux, polymériques, diélectriques, et des colloïdes. Ils présentent des effets de rajeunissement sous l'effet de l'application d'un champ magnétique assez fort, comme les verres peuvent être rajeunis par l'application d'une contrainte. Mais les verres de spin montrent aussi des effets non-triviaux de rajeunissement et mémoire en fonction de la température, effets que certaines études récentes montrent pouvoir aussi exister, quoique moins marqués, dans certains polymères et gels. Nous proposons une description du vieillissement des systèmes désordonnés en termes de combinaison de tels effets, « sélectifs » en température, avec les effets plus habituels, « cumulatifs », liés à la vitesse de refroidissement.*

## 1. Vieillissement dans les verres de spin

Les verres de spin sont des matériaux constitués d'atomes magnétiques entre lesquels les interactions sont aléatoirement ferro- ou antiferro-magnétiques. La situation la plus connue est celle d'alliages inter-métalliques, obtenus par la dilution de quelques % d'atomes magnétiques dans une matrice métallique non-magnétique ; d'autres verres de spin, aux propriétés strictement équivalentes, sont obtenus par dilution de composés magnétiques isolants. Dans ces systèmes, les moments magnétiques ne peuvent satisfaire simultanément les interactions magnétiques contradictoires auxquelles ils sont soumis par leurs voisins. Cette frustration conduit à l'existence d'une multitude d'états métastables, séparés par des barrières d'énergie de toutes tailles, qui dominent le comportement magnétique des verres de spin en produisant des temps de réponse à toutes les échelles à partir du microscopique ($\sim 10^{-12}$ s), sans limite supérieure observée. Les verres de spin se présentent à nous comme perpétuellement hors d'équilibre (Mydosh, 1993 ; Vincent *et al.,* 1997).

La dynamique lente des verres de spin montre d'intéressantes analogies avec celle des verres structuraux ou polymériques. Une procédure expérimentale standard d'étude de ces propriétés (Fig. 1a) consiste à refroidir le verre de spin jusqu'en dessous de sa température de gel $T_g$ (typiquement vers 0.5-0.9 $T_g$) en présence d'un faible champ magnétique H, à attendre pendant un temps $t_w$ (l'aimantation a une valeur $M_{FC}$ qui reste à très peu près constante), pour ensuite couper ce champ à t=0 et mesurer la relaxation vers zéro de l'aimantation, dite " thermo-rémanente " (TRM). Comme le montre la figure 1a, cette relaxation s'étend sur plusieurs décades de temps (échelle logarithmique), et de plus dépend du temps d'attente $t_w$ : plus $t_w$ est grand, plus la relaxation est lente, le verre de spin " durcit ". Ce phénomène de vieillissement est exactement semblable à celui observé dans la relaxation du module élastique des polymères vitreux (Struik, 1978). Il obéit d'ailleurs aux mêmes lois d'échelles ; les courbes correspondant à 2 valeurs $t_{w1}$ et $t_{w2}$ sont à peu près espacées de $\mu.\log t_{w1}/ t_{w2}$, avec $\mu \leq 1$. Ceci suggère une loi d'échelle en $t/t_w^{\mu}$ ; plus exactement, on utilise $\lambda/t_w^{\mu}$, où $\lambda$ est un temps effectif rendant compte de l'effet du vieillissement au cours de la relaxation elle-même (à temps courts $\lambda \sim t$, voir Struik, 1978). On voit sur la figure 1b que cette variable réduite permet de superposer avec une grande précision les relaxations mesurées pour des $t_w$ variant de 300 à 30000 s (Alba *et al*, 1986).

L'exposant $\mu$ mesuré dans les verres de spin reste toujours inférieur à 1, même dans la limite des champs très faibles. Il diminue lorsque l'amplitude du champ magnétique H (dont la coupure provoque la relaxation) augmente. L'insert de la figure 1b montre un exemple dans lequel μ reste constant (=0.85) pour H variant de 0.001 à 10 Oe, puis diminue pour s'annuler vers H=300 Oe. Lorsque H augmente, μ(H) diminue, c'est à dire que la dépendance en $t_w$ de la relaxation devient plus faible (dans la limite $\mu \rightarrow 0$, $t_w^{\mu}$ ne dépend plus de $t_w$) ; le vieillissement effectué pendant $t_w$ est partiellement ou même complètement effacé par une variation importante de champ magnétique. Ce " rajeunissement " sous l'effet d'une variation de champ est l'exact analogue du rajeunissement par l'application d'une contrainte connu en rhéologie des verres (Struik, 1978), et a été récemment mis en évidence dans des suspensions colloïdales formant des gels et des pâtes (voir par ex. Cloitre *et al*, 2000). Dans tous ces systèmes, l'analyse en loi d'échelle du vieillissement du module élastique ou de cisaillement conduit à un exposant μ qui diminue lorsque la contrainte appliquée pour faire la mesure augmente, de manière comparable au cas présenté en insert de la figure 1b. Dans les colloïdes, pour lesquels la température joue un rôle secondaire, l'application d'une forte contrainte est le point de départ des expériences sur le vieillissement, en remplacement d'un réchauffement au-dessus de $T_g$, dont il n'existe pas d'équivalent.

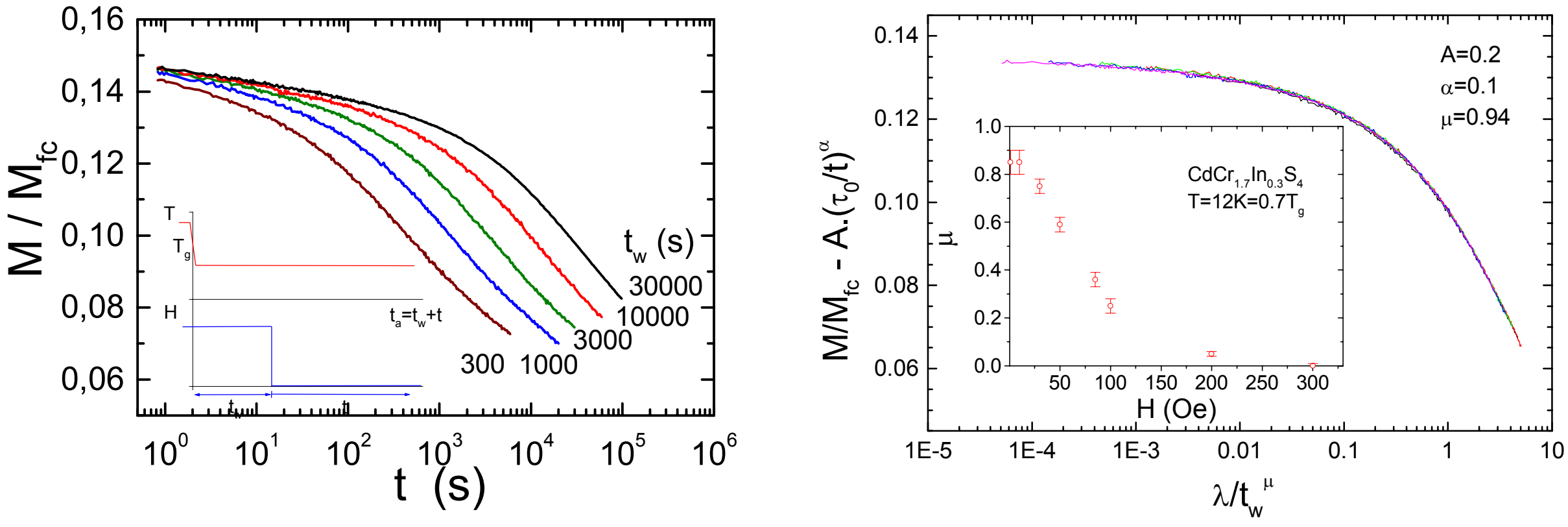

**Figure 1.** (a) Relaxation de l'aimantation thermo-rémanente du verre de spin Ag:$Mn_{2.7\%}$ après coupure de H=0.1 Oe, pour différentes valeurs du temps d'attente $t_w$ avant coupure. (b) Loi d'échelle du vieillissement en $t_w^{\mu}$ : les courbes de (a) se placent sur une courbe maîtresse unique en fonction de la variable réduite $\lambda/t_w^{\mu}$, après soustraction d'une partie indépendante de $t_w$ (Struik 1978, Alba et al 1986, Vincent et al 1996). L'insert montre en exemple la variation de l'exposant μ avec le champ dans le cas d'un autre verre de spin (isolant).

## 2. Effet de la température sur le vieillissement : rajeunissement et mémoire

Dans les verres structuraux et polymériques, la vitesse de refroidissement est un facteur très important. Plus le refroidissement est lent, plus le volume spécifique et l'enthalpie restent proches de leurs valeurs d'équilibre. Des paliers d'attente successifs à différentes températures permettent chacun une relaxation lente vers l'équilibre, et l'effet des différents paliers est considéré, pour l'essentiel, comme cumulatif. Dans le verre de spin, une procédure de ce type met en évidence un comportement assez différent.

La figure 2 montre une expérience dans laquelle on compare l'effet d'un refroidissement rapide et celui d'un refroidissement par paliers sur l'aimantation " ZFC ". Cette aimantation est obtenue, après refroidissement sous champ nul, par l'application d'un petit champ à basse température. L'aimantation atteinte est faible[1] ; on mesure ensuite son augmentation tout en réchauffant progressivement l'échantillon. En fonction de la température, cette courbe montre un pic à $T_g$. Lorsque la même expérience est faite avec 1 ou 2 paliers d'attente de 10000 s à des températures intermédiaires (18 et 12 K) pendant le refroidissement, la valeur de l'aimantation obtenue à basse température après application du champ est la même qu'après un refroidissement direct : les longs temps d'attente au cours desquels l'échantillon a vieilli n'ont aucun effet visible. Par contre, la courbe mesurée en augmentant la température montre des creux très nets autour des températures où ces vieillissements ont été effectués ; leur " mémoire ", stockée au cours du refroidissement par paliers, est révélée au cours du réchauffement. Elle est aussi progressivement effacée au cours de sa lecture en réchauffement (Refregier *et al*, 1987 ; Jonason *et al*, 2000).

[1] L'aimantation ZFC est hors d'équilibre, et relaxe lentement vers $M_{FC}$. Cette procédure est " miroir " de celle de relaxation de la TRM, et donne accès (en champ suffisamment faible) aux mêmes informations ; elle montre en particulier la même dépendance en $t_w$.

Pour un vieillissement effectué à une température $T_1$, la valeur de l'aimantation ZFC à une température $T_2<T_1$ est la même que celle obtenue sans vieillissement à $T_1$ ; de plus, les courbes obtenues avec respectivement 1 palier à $T_2$ et 2 paliers à $T_1$ puis $T_2$ se recouvrent, c'est-à-dire que l'état de vieillissement à $T_2$ est indépendant d'un long vieillissement à une température $T_1$ légèrement supérieure. On peut ici parler de rajeunissement, dans le même sens que plus haut en fonction d'une variation de champ ; il y a eu rajeunissement en baissant la température de $T_1$ à $T_2$, car le vieillissement effectué à $T_1$ parait sans influence sur l'état du système à $T_2$. De nombreuses procédures expérimentales ont été développées dans les verres de spin pour caractériser ces effets de (i) rajeunissement lorsqu'on diminue la température et (ii) mémoire lorsqu'on l'augmente à nouveau (Refregier *et al*, 1987 ; Vincent *et al*, 1996 ; Jonason *et al*, 2000 ; Bouchaud *et al*, 2002). Il est possible d'effectuer une série de vieillissements indépendants en-dessous de $T_g$, et d'obtenir ainsi au réchauffement une courbe présentant 4 ou 5 creux correspondant à ces différentes mémoires.

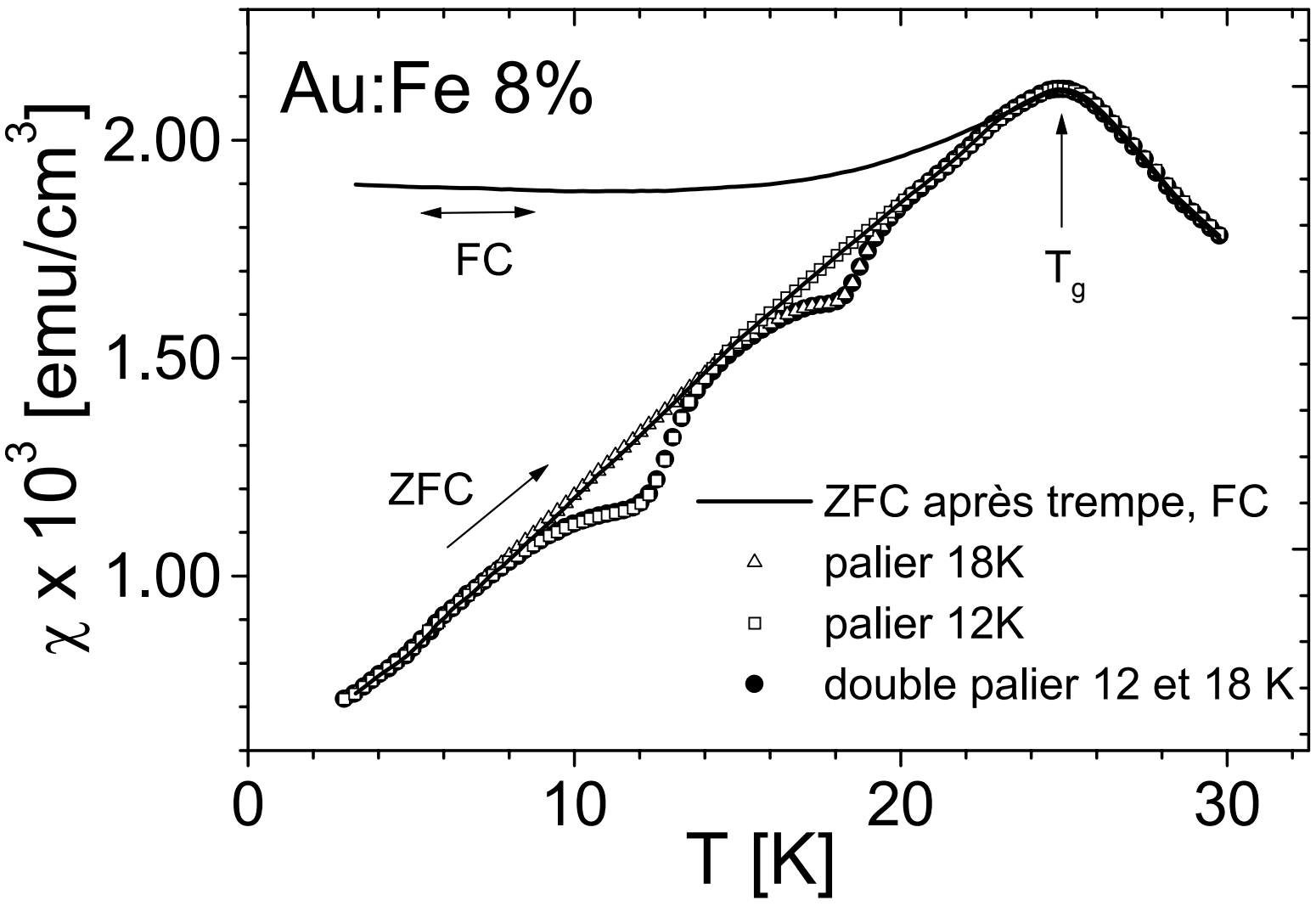

**Figure 2.** *Comparaison des courbes d'aimantation obtenues dans une procédure ZFC par refroidissement direct (trait plein), refroidissement avec 1 palier de 10000s à 18 K (triangles vers le haut), 1 palier à 12 K (triangles vers le bas), et 2 paliers (18 puis 12 K) (cercles).*

## 3. Verres de spin et autres verres

Le verre de spin parait ainsi capable de stocker indépendamment les informations relatives à des vieillissements effectués à différentes températures. Après un refroidissement brutal depuis au-dessus de $T_g$, les spins sont dans une configuration aléatoire, et il est naturel d'imaginer que, au cours du vieillissement, ils s'organisent progressivement pour minimiser leurs énergies d'interaction sur des distances de plus en plus grandes. Cette croissance d'une longueur de corrélation dynamique spin-spin est rendue très lente par la frustration ; elle ne peut pas être facilement imagée comme dans le cas des ferromagnétiques, car le développement de " l'ordre verre de spin " ne correspond à aucune symétrie macroscopique. Les effets de rajeunissement et mémoire ont deux implications importantes sur la croissance d'une telle longueur (Bouchaud *et al*, 2002):

- lorsque la température diminue après un premier palier, le système doit se retrouver hors d'équilibre aux petites échelles de longueur qui vont donner lieu au vieillissement, alors qu'à ces petites échelles il pouvait être à peu près équilibré à plus haute température. Ceci peut se produire du simple fait de la variation thermique des poids de Boltzmann à l'équilibre des diverses configurations ; l'hypothèse d'un " chaos en température ", sans être nécessaire, n'est pas exclue actuellement et donne lieu à diverses discussions (Jonason *et al*, 2000 ; Yoshino *et al*, 2001).
- la possibilité d'effectuer des vieillissements indépendants à plusieurs températures au cours du refroidissement nécessite que les longueurs caractéristiques mises en jeu, de plus en plus petites lorsque la température diminue, varient suffisamment vite avec la température (« hiérarchie » de longueurs en fonction de la température). Dans le cas contraire, le vieillissement peut s'accumuler d'une température à l'autre (ce qui se produit d'ailleurs pour des très petits intervalles de température).

Le verre de spin, dominé par des processus de vieillissement spécifiques de la température, est donc probablement caractérisé par une rapide séparation des échelles de temps et de longueur avec la température (Bouchaud *et al*, 2002). Jusqu'où peut-on considérer que les verres (structuraux et polymériques) doivent leur être opposés en ce que leur vieillissement serait, lui, totalement “ cumulatif ” d'une température à une autre ? Un liquide surfondu tel que le glycérol semble ne pas présenter d'effets de rajeunissement et mémoire (Leheny *et al*, 1998), mais on peut se demander si des mesures avec une procédure plus sensible telle que celle de la Fig. 2 confirmeraient cette conclusion. Des expériences récentes viennent nuancer cette apparente opposition entre verres et verres de spin. Ainsi, la constante diélectrique du PMMA (verre polymérique) garde les traces de vieillissements effectués à des températures différentes[2] ; une double mémoire peut être mise en évidence (Bellon, 2002). Mais la gamme de températures dans laquelle le vieillissement est visible est restreinte, et l'accumulation du vieillissement avec la température est importante ; les creux de mémoire sont donc moins marqués que dans le cas des verres de spin, et s'étalent sur une grande partie de la région où le vieillissement est observable.

Un deuxième exemple vient d'être mis en évidence dans des mesures du module élastique de la gélatine, où des vieillissements spécifiques de la température apparaissent nettement (Parker *et al*, 2002). La gélatine, dont les triples hélices de collagène sont réversiblement dégradables par la température, possède probablement ainsi des degrés de liberté supplémentaires, dont on peut se demander s'ils jouent un rôle dans la possibilité de comportements plus complexes que dans d'autres systèmes vitreux. On peut cependant remarquer que l'essentiel de la phénoménologie des verres structuraux est semblable à celle des polymères, bien que les briques de base des verres (atomes) paraissent a priori plus simples que celles des polymères (chaînes macromoléculaires).

Ce que nous désignons par “ verres ” recouvre une très large gamme de matériaux. Les verres de spin nous donnent un exemple dans lequel des processus de vieillissement spécifiques de la température sont dominants ; dans une moindre mesure, ils présentent également des processus cumulatifs. Les autres verres, quoique dominés par l'accumulation du vieillissement en température, paraissent capables, sur quelques exemples, de processus du type verre de spin. La question de la généralité de l'existence de tels processus dans les matériaux, et de leur interprétation microscopique, est ouverte. L'intérêt des verres de spin pour la compréhension des verres, outre leur simplicité conceptuelle qui facilite en principe la modélisation analytique ou numérique, peut être dans cette grille de déchiffrage qu'ils nous apportent en termes de combinaison de processus de vieillissement plus ou moins fortement spécifiques de la température.

*Nous remercions Jean-Philippe Bouchaud pour son active participation à ce travail.*

## 4. Bibliographie

Alba M., Ocio M., Hammann J., *Europhys. Lett.* **2**, 45 (1986).

Bellon L., Ciliberto S., Laroche C., *Eur. Phys. J. B* **25**, 223 (2002).

Bertin E., Godrèche C., Drouffe J.-M., communication privée, 2001.

Bouchaud J.-P., Dupuis V., Hammann J., Vincent E., *Phys. Rev. B* **65,** 024439 (2002).

Jonason K., Nordblad P., Vincent E., Hammann J., Bouchaud J.-P., *Eur. Phys. J. B* **13**, 99 (2000).

Leheny R.L., Nagel S.R., *Phys. Rev. B* **57**, 5154 (1998).

Mydosh J.A., *Spin glasses : an experimental introduction*, Taylor & Francis, London - Washington 1993.

Parker A., Normand V., à paraître, 2002.

Refregier Ph., Vincent E., Hammann J., Ocio M., *J. Physique (France)* **48**, 1533 (1987).

Struik L.C.E., *Physical aging in amorphous polymers and other materials*, Elsevier, Houston (1978).

Vincent E., Hammann J., Ocio M., Bouchaud J.-P., Cugliandolo L.F., *Complex behaviour of glassy systems*, Springer Verlag Lecture Notes in Physics Vol. **492** pp.184-219, preprint cond-mat/9607224 (1997).

Yoshino H., Lemaître A., Bouchaud J.-P., *Eur. Phys. J. B* **20**, 367 (2001).

[2] L'effet dit de “ mémoire ” ou “ effet Kovacs ” observé dans les polymères après un long vieillissement suivi d'un réchauffement est plus simple que celui évoqué ici, et peut être attendu dans un contexte plus général (Bertin *et al*, 2001).